%% file: main.tex
\documentclass[a4paper, 12pt, numbers=noenddot]{scrartcl}

\input{Template/Preamble}
\input{Template/Settings}

\input{Template/Style}

\begin{document}

\hypersetup{pageanchor=false}
\input{Template/Frontpage}

\hypersetup{pageanchor=true}
\pagenumbering{arabic}
\setcounter{page}{1}
\pagestyle{scrheadings}

\include{Chapters/0.Introduction.tex}
\include{Chapters/1.Stationary_bootstrap.tex}
\include{Chapters/2.VAR_Bootstrap.tex}
\include{Chapters/3.NelsonSiegel.tex}
\include{Chapters/5.Results}

\include{Chapters/8.Results}
\include{Chapters/9.Conclusions}

\appendix
\include{Chapters/A.Appendix}

\newpage
\nocite{*}
\printbibliography[heading=bibintoc]

\end{document}

%% file: Template/Preamble.tex
\usepackage[T1]{fontenc}
\usepackage[utf8]{inputenc}
\usepackage[english]{babel}

\usepackage[table,xcdraw]{xcolor}

\usepackage[left=1.8cm, right=1.8cm, bottom=2.7cm, top=2.7cm]{geometry}
\usepackage[onehalfspacing]{setspace}
\usepackage{ragged2e}
\usepackage{enumitem}
\usepackage{float}
\usepackage[section]{placeins}
\usepackage{verbatim}
\usepackage[headsepline, footsepline]{scrlayer-scrpage}
\usepackage{footnote}
\usepackage[hang,flushmargin]{footmisc}
\usepackage{anyfontsize}

\usepackage{amsmath}
\usepackage{amssymb}
\usepackage{amsthm}
\usepackage{newtxtext,newtxmath}

\theoremstyle{remark}

\theoremstyle{plain}

\usepackage{array}
\usepackage{subcaption}
\usepackage{multirow}
\usepackage{longtable}
\usepackage{tabularx}
\usepackage{makecell}
\usepackage{siunitx}
\usepackage{booktabs}

\usepackage{comment}

\input{Template/Colors}
\usepackage{graphicx}
\usepackage[most]{tcolorbox}
\usepackage{tikz}
\usepackage{pgfplots}
\pgfplotsset{compat=1.18}

\usepackage{csquotes}
\usepackage[backend=biber, style=numeric, sorting=none]{biblatex}

\DefineBibliographyStrings{italian}{%
bibliography={Bibliografia}%
}
\usepackage{xurl}
\usepackage[colorlinks=true, linkcolor=black, citecolor=black, urlcolor=black]{hyperref}

%% file: Template/Colors.tex
\definecolor{arcaBlue}{HTML}{000000}
\definecolor{arcaRed}{HTML}{000000}
\definecolor{arcaWhite}{HTML}{FFFFFF}
\definecolor{arcaBlack}{HTML}{000000}

%% file: Template/Settings.tex
\newcommand{\Title}{Scenario Generation for Time Series and Curves}
\newcommand{\Subtitle}{A Comparison of Nonparametric and Semiparametric Bootstrap}

\newcommand{\ReportDate}{May 2026}

\newcommand{\Keywords}{Stationary Bootstrap; VAR-Bootstrap; Nelson-Siegel; Yield Curve; Scenario Generation; Mean Reversion; Financial Time Series; Term Structure of Interest Rates.}
\newcommand{\JELCodes}{C15; C32; C53; E43; G12; G17.}

\newcommand{\TitlePageAuthors}{%
    \par\noindent
    \makebox[\textwidth][c]{%
        \begin{tabular}{@{}c@{\hspace{0.55cm}}c@{\hspace{0.55cm}}c@{}}
            {\fontsize{13}{16}\selectfont Nicola Baldoni} &
            {\fontsize{13}{16}\selectfont Michele Sparviero} &
            {\fontsize{13}{16}\selectfont Lorenzo Viola} \\[0.35em]
            {\small ARCA Fondi SGR} &
            {\small ARCA Fondi SGR} &
            {\small ARCA Fondi SGR} \\[0.35em]
            {\small\texttt{nicola.baldoni@arcafondi.it}} &
            {\small\texttt{michele.sparviero@arcafondi.it}} &
            {\small\texttt{lorenzo.viola@arcafondi.it}}
        \end{tabular}%
    }%
    \par
}


%% file: Template/Style.tex
\setkomafont{disposition}{\normalfont\bfseries}

\setkomafont{section}{\normalfont\large\bfseries}

\setkomafont{subsection}{\normalfont\normalsize\bfseries}

\setkomafont{subsubsection}{\normalfont\normalsize\itshape}

\setkomafont{paragraph}{\normalfont\normalsize\itshape}
\setkomafont{subparagraph}{\normalfont\normalsize\itshape}

\RedeclareSectionCommand[
  beforeskip=2.5ex plus 1ex minus .2ex,
  afterskip=1.2ex plus .2ex
]{section}

\RedeclareSectionCommand[
  beforeskip=2ex plus .8ex minus .2ex,
  afterskip=0.8ex plus .2ex
]{subsection}

\RedeclareSectionCommand[
  beforeskip=1.5ex plus .6ex minus .2ex,
  afterskip=0.6ex plus .2ex
]{subsubsection}

\clearpairofpagestyles

\setkomafont{pageheadfoot}{\normalfont\small}
\setkomafont{pagenumber}{\normalfont\small}

\newtcolorbox{boxK}{
    colback=white,
    colframe=black,
    boxrule=0.4pt,
    arc=0pt,
    left=6pt,
    right=6pt,
    top=6pt,
    bottom=6pt,
    before skip=1em,
    after skip=1em,
    fontupper=\small
}

%% file: Template/Frontpage.tex
\begin{titlepage}
\thispagestyle{empty}
\begin{singlespace}
\centering

\vspace*{1.3cm}

{\fontsize{20}{25}\selectfont\bfseries \Title\par}

\vspace{0.6cm}

{\fontsize{14}{18}\selectfont \Subtitle\par}

\vspace{1.3cm}

\TitlePageAuthors

\vspace{0.85cm}

{\normalsize \ReportDate\par}

\vspace{1.2cm}

\begin{minipage}{0.82\textwidth}
    \begin{center}
        {\large\bfseries Abstract}
    \end{center}
    \vspace{0.35em}

    {\normalsize\justifying\noindent
        \input{Chapters/00.Astract}\par}

    \vspace{1.0em}

    {\small\raggedright\noindent
        \par}

    \vspace{0.45em}

    {\small\raggedright\noindent
        \textbf{JEL classification:} \JELCodes\par}
\end{minipage}

\vfill
\end{singlespace}
\end{titlepage}

%% file: Chapters/00.Astract.tex
Generating stochastic trajectories for asset classes is an increasingly relevant task in quantitative finance. Traditional approaches, such as the stationary bootstrap, preserve by construction the empirical distribution of asset-class returns, but do not ensure that each individual simulated path is economically realistic: scenarios may be valid in distribution while single trajectories fail to represent plausible states of the world.
To address this limitation, we review semiparametric simulation methodologies that combine a parametric structure, which enforces realistic dynamics, with the resampling of model residuals, which preserves the stochastic component observed in historical data. The issue is particularly acute for interest rates, where direct resampling of rate changes may produce implausible yield-curve evolutions despite correct distributional properties. Our empirical analysis shows the effectiveness of semiparametric bootstrap methods based on autoregressive or mean-reverting specifications. In the fixed-income setting, combining these methods with fully parametric term-structure models yields more coherent and realistic simulations of yield-curve dynamics.

%% file: Chapters/0.Introduction.tex
\section{Introduction}
This report analyses and compares several simulation methodologies for multivariate financial time series and yield curves. The objective is to assess the extent to which approaches based on empirical resampling and models with an explicit dynamic structure are capable of generating scenarios consistent with the characteristics observed in historical data. The analysis considers both factors represented in terms of price levels, such as equity indices, and variables modelled in additive terms, such as interest rates and inflation.
Section \ref{sec:stationary_bootstrap} introduces the Stationary Bootstrap, which represents our standard methodology. This approach allows certain empirical properties of historical series to be preserved through block resampling, in a fully nonparametric fashion. Section \ref{sec:var_bootstrap} then presents the VAR-Bootstrap extension, in which the nonparametric resampling of residuals is combined with a vector autoregressive component. This formulation makes it possible to incorporate an explicit dependence on the current level of the variables and, in particular for interest rate factors, to represent potential \emph{mean reversion} dynamics.
Section \ref{sec:nelson_siegel} is devoted to the Nelson-Siegel model, which is employed to provide a parsimonious description of the term structure of interest rates by means of a small number of interpretable factors. Subsequently, in Section \ref{sec:ns_var_bootstrap}, this representation is integrated into the VAR-Bootstrap procedure, so as to model the dynamics of the yield curve through the evolution of latent factors rather than through the individual rates observed at the various maturities.
Finally, Section \ref{sec:results} presents the empirical comparison among the methodologies considered. The results are evaluated with respect to complementary metrics, designed to measure the adherence of the simulated distributions to the historical ones, the ability to reproduce the dependence structure among the factors, and the financial plausibility of the generated curves. In particular, the analysis highlights the trade-off between fidelity to the empirical distribution and the introduction of dynamic or structural constraints, with particular attention to the regularity and consistency of the simulated yield curves.

%% file: Chapters/1.Stationary_bootstrap.tex
\section{Stationary Bootstrap}
\label{sec:stationary_bootstrap}

The baseline methodology adopted for the simulation of financial time series is the Stationary Bootstrap, following the framework of Politis and Romano \cite{politis1994stationary}. In what follows, this procedure constitutes the reference method against which the extensions proposed in this work are compared. For ease of exposition, the construction is described with reference to a single simulated trajectory; the extension to a set of simulations is straightforward and is not made explicit, solely in order to avoid burdening the notation.

Consider \(n_x\) factors represented in terms of price levels. Let
\[
\mathbf{P}_t
=
\bigl(P_t^{(1)},\dots,P_t^{(n_x)}\bigr)^\top
\in \mathbb{R}^{n_x}
\]
denote the vector of levels observed at time \(t\). Such variables are characterised by a non-negativity constraint, namely \(P_t^{(i)}\geq 0\) for every component \(i=1,\dots,n_x\), and are therefore naturally described through relative variations. This class includes, for example, equity prices and market indices. Consistently with this setting, the dynamics of the price factors are reduced to the vector of returns
\[
\mathbf{r}_t
=
\bigl(r_t^{(1)},\dots,r_t^{(n_x)}\bigr)^\top,
\qquad
r_t^{(i)}
=
\frac{P_t^{(i)}-P_{t-1}^{(i)}}{P_{t-1}^{(i)}},
\qquad i=1,\dots,n_x.
\]
Resampling is therefore performed on the returns, so as to represent the evolution of the levels in proportional terms and to preserve the multiplicative nature of the dynamics of the price factors.

In addition, consider \(n_y\) rate factors, whose evolution is described in additive terms, and let
\[
\mathbf{y}_t
=
\bigl(y_t^{(1)},\dots,y_t^{(n_y)}\bigr)^\top
\in \mathbb{R}^{n_y}
\]
denote the corresponding vector observed at time \(t\). Unlike price levels, these variables are not subject to a positivity constraint and may therefore also take negative values. This category includes, for example, interest rates and inflation rates. Consistently with this different nature, the dynamics of such factors are described by means of absolute variations, defined as
\[
\Delta \mathbf{y}_t
=
\bigl(\Delta y_t^{(1)},\dots,\Delta y_t^{(n_y)}\bigr)^\top,
\qquad
\Delta y_t^{(j)}
=
y_t^{(j)}-y_{t-1}^{(j)},
\qquad j=1,\dots,n_y.
\]
Resampling is therefore performed on the first differences, so as not to introduce sign constraints on the simulated values.

The multidimensional structure of the problem is thus represented by the joint vector
\[
\mathbf{z}_t
=
\begin{pmatrix}
\mathbf{r}_t\\
\Delta \mathbf{y}_t
\end{pmatrix}
\in \mathbb{R}^{n},
\qquad n=n_x+n_y.
\]
In the Stationary Bootstrap, resampling is carried out by means of consecutive blocks drawn from the historical sequence \(\{\mathbf{z}_t\}\). This yields a bootstrap trajectory
\[
\mathbf{z}_1^{*},\mathbf{z}_2^{*}, \dots , \mathbf{z}_t^{*}, \dots ,\mathbf{z}_T^{*} 
\qquad
\mathbf{z}_t^{*}
=
\begin{pmatrix}
\mathbf{r}_t^{*}\\
\Delta \mathbf{y}_t^{*}
\end{pmatrix},
\]
where \(\mathbf{r}_t^{*}\in\mathbb{R}^{n_x}\) and \(\Delta \mathbf{y}_t^{*}\in\mathbb{R}^{n_y}\) denote, respectively, the resampled returns and the resampled variations. For the constructive details of the procedure, the reader is referred to Politis and Romano \cite{politis1994stationary}.

The simulation of the price factors is then defined by
\[
\widehat{\mathbf{P}}_t
=
\bigl(\mathbf{1}_{n_x}+\mathbf{r}_t^{*}\bigr)\odot \widehat{\mathbf{P}}_{t-1},
\]
where \(\mathbf{1}_{n_x}\) is the unit vector of dimension \(n_x\) and \(\odot\) denotes the element-wise product. Equivalently, for each component \(i=1,\dots,n_x\),
\[
\widehat{P}_t^{(i)}
=
\bigl(1+(r_t^{(i)})^{*}\bigr)\widehat{P}_{t-1}^{(i)}.
\]

Analogously, the simulation of the rate factors is given by
\[
\widehat{\mathbf{y}}_t
=
\widehat{\mathbf{y}}_{t-1}+\Delta \mathbf{y}_t^{*},
\]
that is, for each component \(j=1,\dots,n_y\),
\[
\widehat{y}_t^{(j)}
=
\widehat{y}_{t-1}^{(j)}+(\Delta y_t^{(j)})^{*}.
\] 

The defining property of the Stationary Bootstrap concerns its ability to
reproduce, in a distributional and asymptotic sense, the law of the underlying
process from the observed sample alone. Let \(\widehat F_T\) denote the empirical
joint distribution induced by \((\mathbf z_1,\ldots,\mathbf z_T)\) and \(F\) the
corresponding law of the data-generating process. Assume \(\{\mathbf z_t\}\) is
strictly stationary. Then, under suitable assumptions, the bootstrap sample \((\mathbf z_1^*,\ldots,\mathbf z_T^*)\)
generated by geometric-block resampling satisfies
\[
\widehat F_T^{\,*}\overset{\mathbb P}{\to} F
\] where \(\widehat F_T^{\,*}\) is the empirical distribution of the bootstrap path
and convergence is understood for the finite-dimensional distributions. The
replication therefore occurs at the level of the probability measure governing the
ensemble of trajectories, not at the level of any individual path.

For a statistic \(\widehat\theta_T=\theta(\mathbf z_1,\ldots,\mathbf z_T)\) admitting
a non-degenerate limit under normalisation \(a_T>0\), the same conditions yield
\[
d\!\left(
\mathcal L^*\!\left(a_T(\widehat\theta_T^*-\widehat\theta_T)\right),\;
\mathcal L\!\left(a_T(\widehat\theta_T-\theta_0)\right)
\right)
\xrightarrow{\mathbb P}0,
\] with \(\mathcal L^*(\,\cdot\,)\) denoting the conditional law of its argument given
the observed sample \((\mathbf z_1,\ldots,\mathbf z_T)\) under the bootstrap
resampling measure \(\mathbb P^*\), \(\mathcal L(\,\cdot\,)\) the unconditional law
of its argument under the data-generating measure \(\mathbb P\), and \(d\) any
metric inducing weak convergence on the space of Borel probability measures on
\(\mathbb R\).

A direct consequence is that the Stationary Bootstrap imposes no equality between
the sample moments of any individual bootstrap trajectory and the corresponding
historical moments. Means, volatilities and correlations computed along a finite
bootstrap path remain random quantities, and their agreement with the empirical
counterparts holds only asymptotically and distributionally for the statistic
under consideration, not as a pathwise matching constraint.
The Stationary Bootstrap thus provides a nonparametric representation of the joint distribution of the risk factors, constructed on the basis of the empirical evidence drawn from the historical sample. It nonetheless retains the limitation, common to purely empirical resampling methods, of not explicitly incorporating specific long-run dynamic properties, such as \emph{mean reversion}, which may be relevant for certain classes of financial assets.

%% file: Chapters/2.VAR_Bootstrap.tex
\section{VAR-Bootstrap}
\label{sec:var_bootstrap}

For certain classes of financial factors, and in particular for interest rates, a simulation constructed exclusively on the resampling of observed variations may prove insufficient to describe their medium- to long-term dynamics. Indeed, the current level of the variable is itself economically relevant information: particularly high or particularly low values are unlikely to be compatible, over long horizons, with an indefinite continuation of the local dynamics implicit in the differences alone. In such a context, it is therefore appropriate to adopt a specification in which future evolution depends not only on past variations, but also on the level reached by the process, so as to be able to incorporate a \emph{mean reversion} mechanism. It is worth noting that the introduction of mean reversion dynamics through an autoregressive specification constitutes an explicit parametric assumption. This assumption nevertheless remains relatively weak, since it does not impose \emph{ex ante} the presence of an actual pull towards the mean: the intensity of such a mechanism is in fact determined by the estimated parameters and may, in principle, turn out to be zero. In this sense, the model does not impose mean reversion dynamics, but allows them to be represented whenever they find support in the data.

To this end, a VAR-Bootstrap model is considered. This construction combines a vector autoregressive specification with a nonparametric resampling of the residuals. Here too, for ease of exposition, the procedure is described with reference to a single simulated trajectory; the extension to a set of simulations is straightforward and is omitted in order not to burden the notation.

Consider \(n_x\) price factors and \(n_y\) rate factors, and define \(n=n_x+n_y\). We introduce the state vector
\[
\mathbf{x}_t=
\begin{pmatrix}
\mathbf{r}_t\\
\mathbf{y}_t
\end{pmatrix}
\in\mathbb{R}^{n},
\]
where \(\mathbf{r}_t\in\mathbb{R}^{n_x}\) is the vector of returns of the price factors at time \(t\), while \(\mathbf{y}_t\in\mathbb{R}^{n_y}\) is the vector of levels of the rate factors. In this formulation, the price component is therefore represented in terms of returns, while the rate component is represented in terms of levels.

The VAR(1) model is defined as
\begin{equation}
    \mathbf{x}_t=\mathbf{a}_0+A_1\mathbf{x}_{t-1}+\boldsymbol{\eta}_t,
    \label{eq:VAR}
\end{equation}
where \(\mathbf{a}_0\in\mathbb{R}^{n}\), \(A_1\in\mathbb{R}^{n\times n}\) and \(\boldsymbol{\eta}_t\in\mathbb{R}^{n}\) is a vector of zero-mean residuals.

Once the model parameters have been estimated on the historical sample, the corresponding residuals are defined as
\begin{equation}
    \widehat{\boldsymbol{\eta}}_t
    =
    \mathbf{x}_t-\widehat{\mathbf{a}}_0-\widehat{A}_1\mathbf{x}_{t-1}.
\end{equation}

The simulation is then constructed by means of
\begin{equation}
    \widehat{\mathbf{x}}_t
    =
    \widehat{\mathbf{a}}_0+\widehat{A}_1\widehat{\mathbf{x}}_{t-1}
    +\boldsymbol{\eta}_t^{*},
    \label{eq:var_bootstrap_sim}
\end{equation}
where \(\boldsymbol{\eta}_t^{*}\) denotes an innovation obtained by resampling from the historical residuals \(\{\widehat{\boldsymbol{\eta}}_t\}\), in a manner analogous to the empirical resampling applied to increments in the Stationary Bootstrap. In this way, the deterministic component of the dynamics is described by the VAR(1) model, while the distribution of the residuals is treated in nonparametric form.

The \emph{mean reversion} property is implicit in the autoregressive structure of the model. In particular, if the spectral radius\footnote{The spectral radius of a matrix is the maximum, in absolute value, of its eigenvalues.} of the matrix \(A_1\), denoted by \(\rho(A_1)\), satisfies the condition
\begin{equation}
    \rho(A_1)<1,
\end{equation}
the process is stationary and admits an invariant mean equal to
\begin{equation*}
    \boldsymbol{\mu}
    =
    (I-A_1)^{-1}\mathbf{a}_0.
\end{equation*}
In such a configuration, the matrix \(A_1\) determines the way in which deviations from the mean value propagate over time. The condition on the spectral radius ensures that such deviations are not amplified indefinitely, but tend progressively to dampen. The autoregressive dynamics therefore introduces a pull towards the long-run level \(\boldsymbol{\mu}\), the value of which depends on the estimated coefficients of the model.

Once the simulated trajectory \(\{\widehat{\mathbf{x}}_t\}\) has been obtained, the equity component in levels is reconstructed from the simulated returns. If \(\widehat{\mathbf{r}}_t\) denotes the equity subcomponent of \(\widehat{\mathbf{x}}_t\), one sets
\[
\widehat{\mathbf{P}}_t
=
\bigl(\mathbf{1}_{n_x}+\widehat{\mathbf{r}}_t\bigr)\odot \widehat{\mathbf{P}}_{t-1},
\]
where \(\widehat{\mathbf{P}}_t\in\mathbb{R}^{n_x}\) is the vector of simulated levels of the equity factors. The rate component, being modelled directly in levels, is instead obtained as the subvector \(\widehat{\mathbf{y}}_t\) of \(\widehat{\mathbf{x}}_t\).

A limitation of the procedure lies in the growth of the number of parameters with the dimension of the system. Indeed, the matrix \(A_1\) contains \(n^2\) coefficients, with potential issues both in terms of robustness of the estimation and in terms of computational burden when the number of factors is large. This aspect can be partially mitigated by means of a reduced-dimension representation of the term structure of interest rates, for instance through Nelson-Siegel-type latent factors.

%% file: Chapters/3.NelsonSiegel.tex
\section{Nelson-Siegel Model}
\label{sec:nelson_siegel}

The Nelson-Siegel model \cite{nelson1987parsimonious} makes it possible to represent the term structure of interest rates in a parsimonious form, synthesising the information contained in the curve by means of a reduced number of factors. Such a representation proves particularly useful when the observed curve is available over a wide range of maturities, since it allows a pointwise description of the rates to be replaced by a low-dimensional parametric specification.

Let \(y_t(\tau)\) denote the interest rate observed at time \(t\) and corresponding to maturity \(\tau\). The dynamic parameterisation of Nelson-Siegel, following the framework of Diebold and Li \cite{diebold2006forecasting}, can be expressed as
\begin{equation}
    y_t(\tau)
    =
    \beta_{0,t}
    +
    \beta_{1,t}L_1(\tau)
    +
    \beta_{2,t}L_2(\tau).
    \label{eq:nelson_siegel}
\end{equation}

With respect to the original formulation, in which the coefficients are constant over time, the parameters \(\beta_{0,t}\), \(\beta_{1,t}\) and \(\beta_{2,t}\) are here interpreted as dynamic latent factors. For each date \(t\), they synthesise, respectively, the level, slope, and curvature components of the observed curve; along the time dimension, they instead describe the evolution of the principal movements of the term structure.

The functions \(L_1(\tau)\) and \(L_2(\tau)\) represent the factor loadings associated with the slope and curvature components, determining the way in which such factors affect the various maturities. They are defined as
\begin{equation}
\label{eq:carichi}
\begin{split}
    L_1(\tau)
    =
    \frac{1-e^{-\lambda \tau}}{\lambda \tau}, \qquad
    L_2(\tau)
    =
    \frac{1-e^{-\lambda \tau}}{\lambda \tau}
    -
    e^{-\lambda \tau}.
\end{split}
\end{equation}

\begin{figure}[H] 
    \centering
    \begin{minipage}{0.333\textwidth}
        \centering
        \includegraphics[width=\linewidth]{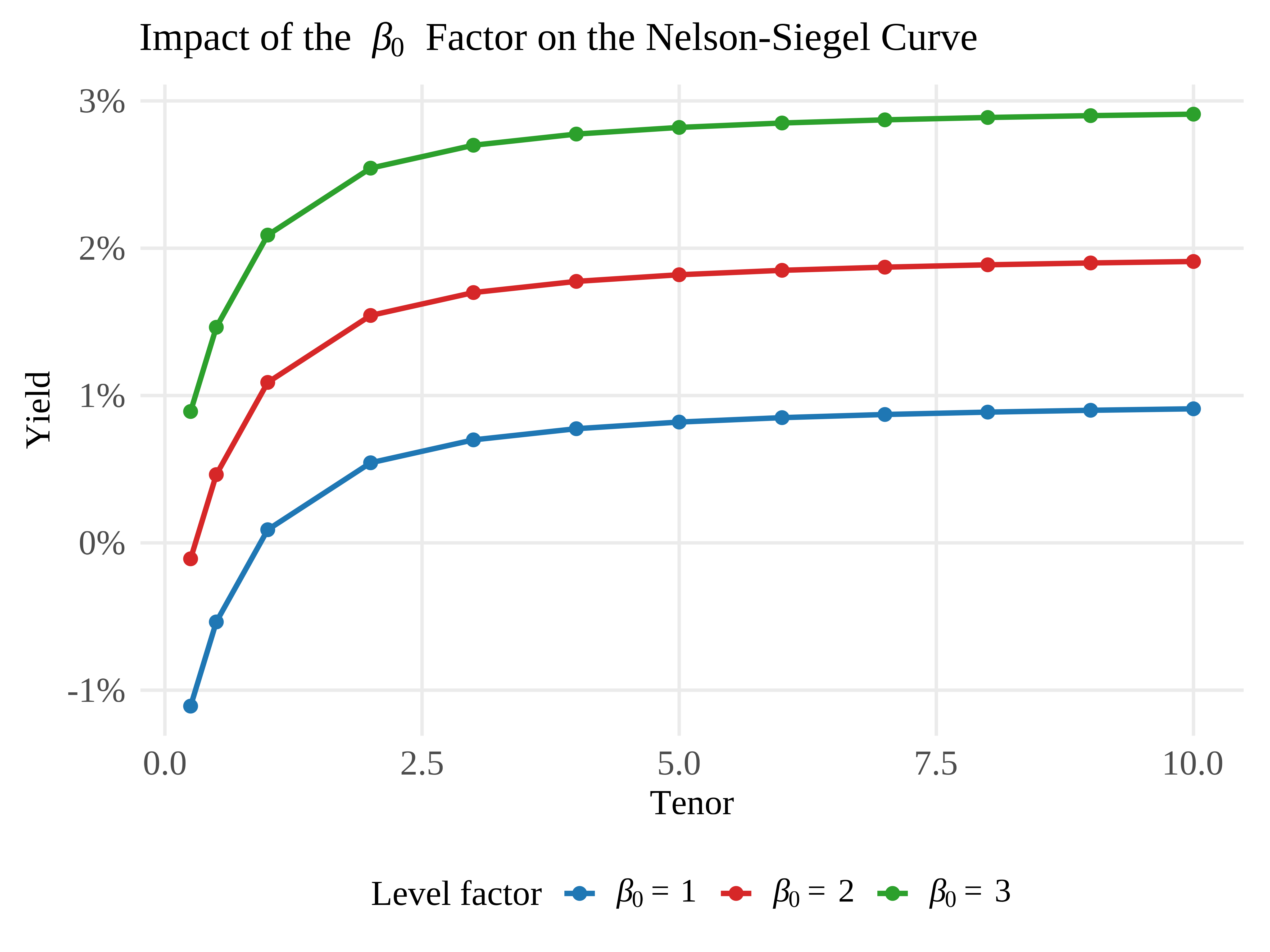}
    \end{minipage}%
    \hfill%
    \begin{minipage}{0.333\textwidth}
        \centering
        \includegraphics[width=\linewidth]{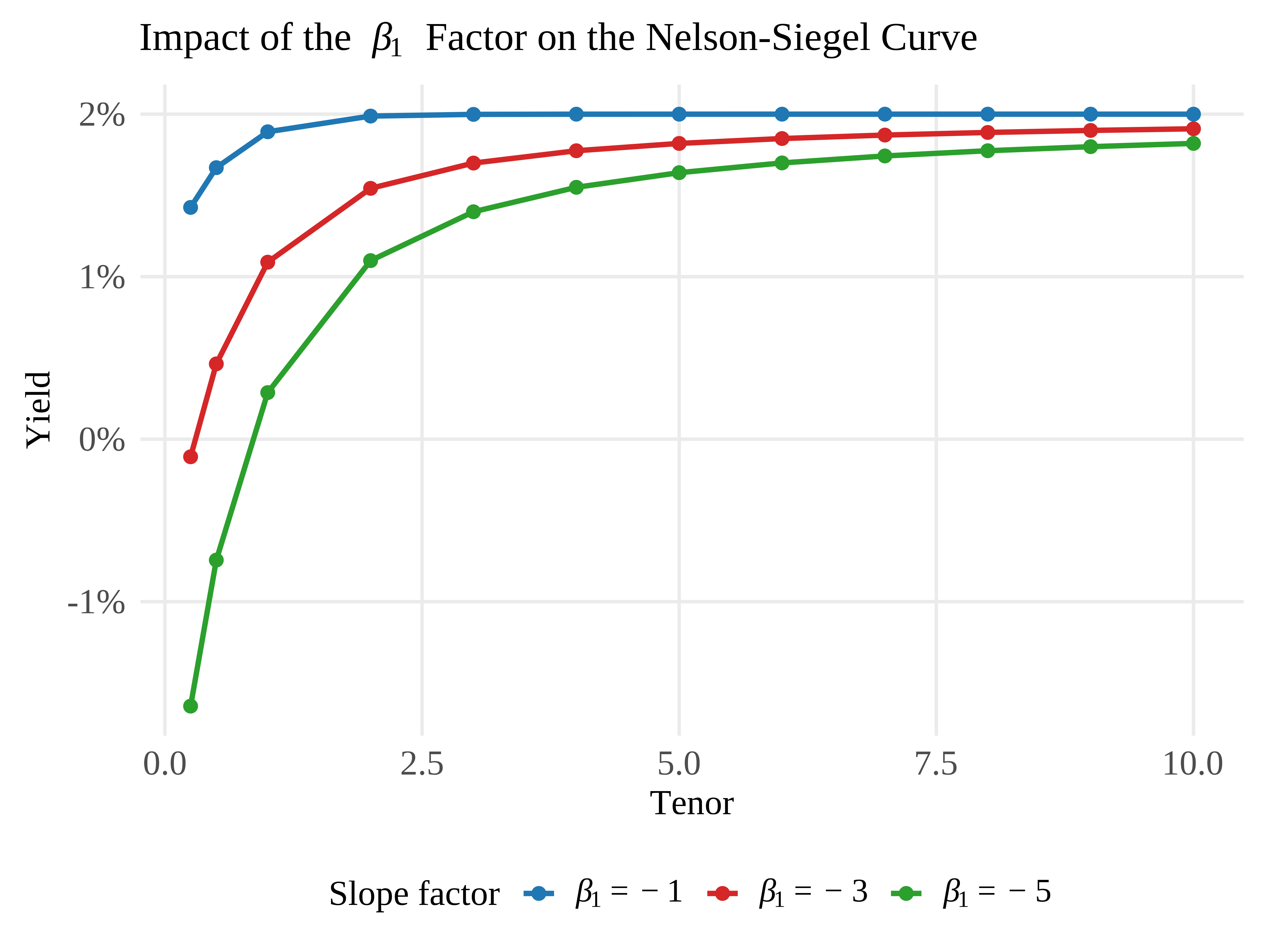}
    \end{minipage}%
    \hfill%
    \begin{minipage}{0.333\textwidth}
        \centering
        \includegraphics[width=\linewidth]{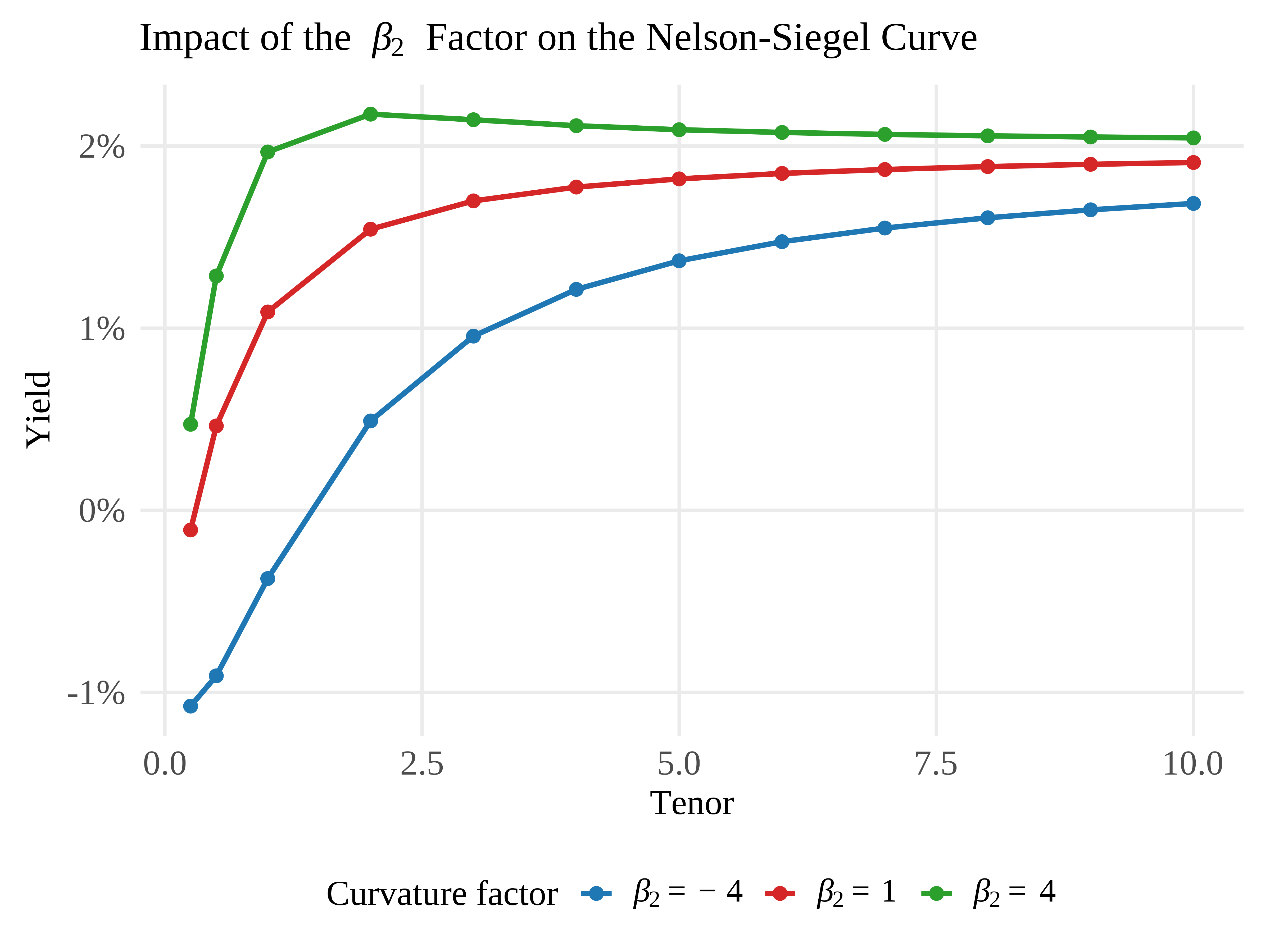}
    \end{minipage}
    \caption{Effect of the Nelson-Siegel factors on the term structure of interest rates. Each panel shows the impact of the variation of a single parameter, attributable respectively to the level, slope, and curvature components. The red curve is obtained with the same parametric configuration in the three plots and therefore constitutes the reference profile.}
    \label{fig:ns_beta}
\end{figure}

In particular, the level factor acts uniformly on the curve, the slope factor mainly affects short maturities, while the curvature factor allows variations that are more localised at intermediate maturities to be represented. Figure \ref{fig:ns_beta} separately illustrates the effect of the three factors on the shape of the term structure.

The parameter \(\lambda\) regulates the speed at which the factor loadings \(L_1(\tau)\) and \(L_2(\tau)\) vary as the maturity increases. It therefore determines the region of the term structure in which the curvature component exerts its principal effect. Higher values of \(\lambda\) concentrate such an effect on short maturities, while lower values shift its incidence towards longer maturities.

In the present treatment, the parameter \(\lambda\) is regarded as exogenous. This choice makes it possible to separate the calibration of the decay parameter from the estimation of the Nelson-Siegel factors. In particular, once \(\lambda\) has been fixed, the coefficients \(\beta_{0,t}\), \(\beta_{1,t}\) and \(\beta_{2,t}\) can be estimated, for each observation date, by means of a least-squares optimisation procedure. One thus obtains a time series of Nelson-Siegel factors, which constitutes a reduced-dimension representation of the dynamics of the yield curve. The value of \(\lambda\) is selected by minimising the mean squared error between the observed rates and those reconstructed by the model over the entire historical series. The calibration is carried out by considering only maturities less than or equal to five years, consistently with the horizon of interest of the analysis.
\footnote{
A possible generalisation of the Nelson-Siegel model is represented by the Svensson model, which introduces an additional curvature factor and a second decay parameter. Such an extension allows greater flexibility in the representation of the term structure, in particular in cases in which the curve exhibits more than one relevant inflection. In the present work, however, the Svensson model is not adopted. The increase in flexibility offered by such a specification in fact entails an increase in the number of parameters to be estimated and, consequently, a greater complexity of the calibration procedure. In light of the objectives of the analysis, it has been preferred to retain the Nelson-Siegel formulation in the dynamic variant of Diebold and Li, which provides a parsimonious, interpretable, and temporally evolving representation of the yield curve. Recourse to the Svensson model nonetheless remains a possible extension for future developments.}

\section{Nelson-Siegel VAR-Bootstrap}
\label{sec:ns_var_bootstrap}

In what follows, the dynamic Nelson-Siegel representation is integrated into the VAR-Bootstrap procedure in order to obtain a reduced-dimension simulation of the yield curve. The underlying idea consists in replacing the direct modelling of the rates observed at the individual maturities with the modelling of the latent factors that govern their shape: $\beta_0,\ \beta_1,\ \beta_2$. In this way, the autoregressive dynamics is not applied separately to each node of the curve, but rather to the level, slope, and curvature factors, from which the simulated curve is subsequently reconstructed.

In the presence of additional financial factors beyond the yield curve, the VAR-Bootstrap is applied to the joint state vector
\begin{equation*}
    \mathbf{x}_t
    =
    \begin{pmatrix}
    \mathbf{r}_t\\
    \mathbf{y}_t\\
    \boldsymbol{\beta}_t
    \end{pmatrix},
\end{equation*}
where \(\mathbf{r}_t\) represents the returns of the price factors, \(\mathbf{y}_t\) collects any factors modelled in levels and not belonging to the interest rate curve, while \(\boldsymbol{\beta}_t\) contains the Nelson-Siegel factors. For ease of exposition, in what follows the notation is developed explicitly only with reference to the component \(\boldsymbol{\beta}_t\), since this is the part relevant for the reconstruction of the term structure. Moreover, in order not to burden the notation, the procedure is described with reference to a single simulated trajectory; the extension to a set of simulations is straightforward.

Let \(y_t(\tau)\) denote the rate observed at time \(t\) for maturity \(\tau\). In the dynamic setting of Diebold and Li, the yield curve is represented as
\begin{equation}
    y_t(\tau)
    =
    \beta_{0,t}
    +
    \beta_{1,t}L_1(\tau)
    +
    \beta_{2,t}L_2(\tau)
    +
    \varepsilon_t(\tau),
    \label{eq:tassi_NS}
\end{equation}
where \(\varepsilon_t(\tau)\) represents the fitting error of the model at the specific maturity, while the values of $L_1(\tau)$ and $L_2(\tau)$ are defined in equation \eqref{eq:carichi}.

The parameter \(\lambda\) is fixed according to the calibration criterion illustrated in Section \ref{sec:nelson_siegel}. Conditional on this value, the coefficients \(\beta_{0,t}\), \(\beta_{1,t}\) and \(\beta_{2,t}\) are estimated, for each date, by means of quadratic optimisation. One thus obtains the historical series of the vectors
\begin{equation*}
    \boldsymbol{\beta}_t
    =
    \bigl(\beta_{0,t},\beta_{1,t},\beta_{2,t}\bigr)^\top,
\end{equation*}
which synthesise the dynamics of the yield curve in three latent factors.

The temporal dynamics of the factors is described by means of a VAR(1) model:
\begin{equation}
    \boldsymbol{\beta}_t
    =
    \mathbf{a}_0
    +
    A_1\boldsymbol{\beta}_{t-1}
    +
    \boldsymbol{\eta}_t,
    \label{eq:VAR_NS}
\end{equation}
where \(\mathbf{a}_0\in\mathbb{R}^3\), \(A_1\in\mathbb{R}^{3\times 3}\) and \(\boldsymbol{\eta}_t\in\mathbb{R}^3\) is the vector of innovations. After the estimation of the parameters, the residuals of the model are resampled with replacement and used to generate the simulated trajectory of the factors:
\begin{equation}
    \widehat{\boldsymbol{\beta}}_t
    =
    \widehat{\mathbf{a}}_0
    +
    \widehat{A}_1\widehat{\boldsymbol{\beta}}_{t-1}
    +
    \boldsymbol{\eta}_t^{*}.
    \label{eq:VAR_NS_bootstrap}
\end{equation}
The autoregressive component thus describes the dynamics of the latent factors, while the resampling of the residuals allows the distribution to be treated in nonparametric form.

Once the simulated factors have been obtained, the yield curve is reconstructed by again applying the Nelson-Siegel relation. Defining
\begin{equation*}
    \boldsymbol{\ell}(\tau)
    =
    \bigl(1,L_1(\tau),L_2(\tau)\bigr),
\end{equation*}
the simulated rate at maturity \(\tau\) is given by
\begin{equation}
    \widehat{y}_t(\tau)
    =
    \boldsymbol{\ell}(\tau)\widehat{\boldsymbol{\beta}}_t
    =
    \boldsymbol{\ell}(\tau)\widehat{\mathbf{a}}_0
    +
    \boldsymbol{\ell}(\tau)\widehat{A}_1\widehat{\boldsymbol{\beta}}_{t-1}
    +
    \boldsymbol{\ell}(\tau)\boldsymbol{\eta}_t^{*}.
    \label{eq:tassi_NS_sim}
\end{equation}

It should be emphasised that the simulation concerns the factor component of the curve. The term \(\varepsilon_t(\tau)\), which measures the fitting error of the Nelson-Siegel specification at the individual maturity, is not resampled. Consequently, the procedure reproduces the dynamics of the portion of the curve explained by the latent factors, but does not incorporate the residual variability specific to the individual nodes. An analysis of the residuals $\varepsilon_t(\tau)$ is presented in Section \ref{sec:results_curve}.

%% file: Chapters/5.Results.tex
\section{Results}
\label{sec:results}

This section proceeds with the comparison among the different simulation methodologies considered. The assessment is carried out by means of a set of metrics aimed at capturing complementary aspects of the simulated trajectories: the incidence and intensity of negative increments, the dependence structure among the factors, and the consistency of the simulated distributions with respect to those historically observed.

The empirical sample used in the analysis comprises three time series representative of the main classes of factors considered. The equity component is described by the MSCI Europe Net Total Return EUR Index, hereinafter referred to as Equity EUR. The rate component is represented by the gross yield of 12-month Italian Treasury Bills (BOT), used as a proxy for the 1-year Italian yield. Finally, the inflation component is obtained from the Harmonised Index of Consumer Prices of the Euro area, excluding tobacco, in its non-revised and non-seasonally adjusted version.

For each methodology, \(10^4\) trajectories are simulated over a time horizon of 5 years. In the case of daily series, this horizon corresponds to \(1260\) time nodes; in the case of monthly series, it corresponds instead to \(60\) observations.

The comparison first examines the incidence and intensity of negative increments. By this expression are meant, depending on the nature of the variable considered, the negative returns of the factors represented in terms of prices and the negative variations of the factors modelled in additive terms. The percentage of negative increments therefore measures the frequency with which the simulation generates downward movements with respect to the previous period. The mean value of the negative increments, computed conditionally on the episodes in which the increment is below zero, provides instead a measure of the average intensity of such movements. These metrics make it possible to assess whether the various models reproduce in a consistent manner not only the frequency, but also the average magnitude of the negative variations observed in the historical sample.

The correlation among the simulated variables is used to verify the extent to which each methodology is capable of preserving the dependence structure observed in the historical sample. This aspect is central in a multivariate context, since a correct representation of the relationships among the risk factors is necessary in order to obtain financially plausible joint scenarios.

Finally, the Kullback--Leibler divergence is employed to compare the historical empirical distribution with that obtained from the simulations. It is defined as
\begin{equation}
    D_{KL}(f_1 \| f_2)
    :=
    \int f_1(x)\,
    \log\!\left(
    \frac{f_1(x)}{f_2(x)}
    \right)
    \, dx,
    \label{eq:KL}
\end{equation}
where \(f_1\) denotes the density estimated on the historical data and \(f_2\) the density estimated on the simulated data. The divergence measures the informational loss associated with the use of the simulated distribution to approximate the historical one. Smaller values therefore indicate greater adherence of the simulation to the observed empirical distribution. Since the densities are not known in closed form, they are estimated numerically by means of \emph{kernel density estimation}; consequently, the value of the metric also depends on the choice of the \emph{bandwidth} parameter, which regulates the degree of smoothing of the estimate.

It should be noted that such metrics are not to be interpreted as equivalent criteria. If the objective were exclusively the reproduction of the historical distribution, a purely empirical resampling procedure would naturally be at an advantage, since it generates observations directly from the available sample. The VAR-Bootstrap and Nelson-Siegel VAR-Bootstrap models, by contrast, introduce a parametric dynamic structure, with the aim of incorporating properties such as temporal dependence, mean reversion, and a parsimonious representation of the yield curve. The comparison should therefore be read as an assessment of the trade-off between statistical adherence to the historical data and economic-financial plausibility of the simulated scenarios.

\section{Comparison between Stationary Bootstrap and VAR-Bootstrap}

In light of these considerations, the empirical comparison is organised into two specifications. The objective is to separate the effect of data frequency and the inclusion of inflation.

The first specification compares the Stationary Bootstrap and the VAR-Bootstrap on two factors, Equity EUR and the 1-year Italian rate, using daily observations. The second extends the system by also including inflation; since the latter is available only at monthly frequency, the simulation is in this case carried out on monthly data. For the inflation component, dynamics analogous to those of the rate factors are assumed, with its long-run level constrained to \(2\%\).

\begin{figure}
    \centering

    \begin{minipage}{0.499\textwidth}
        \centering
        \includegraphics[width=\linewidth]{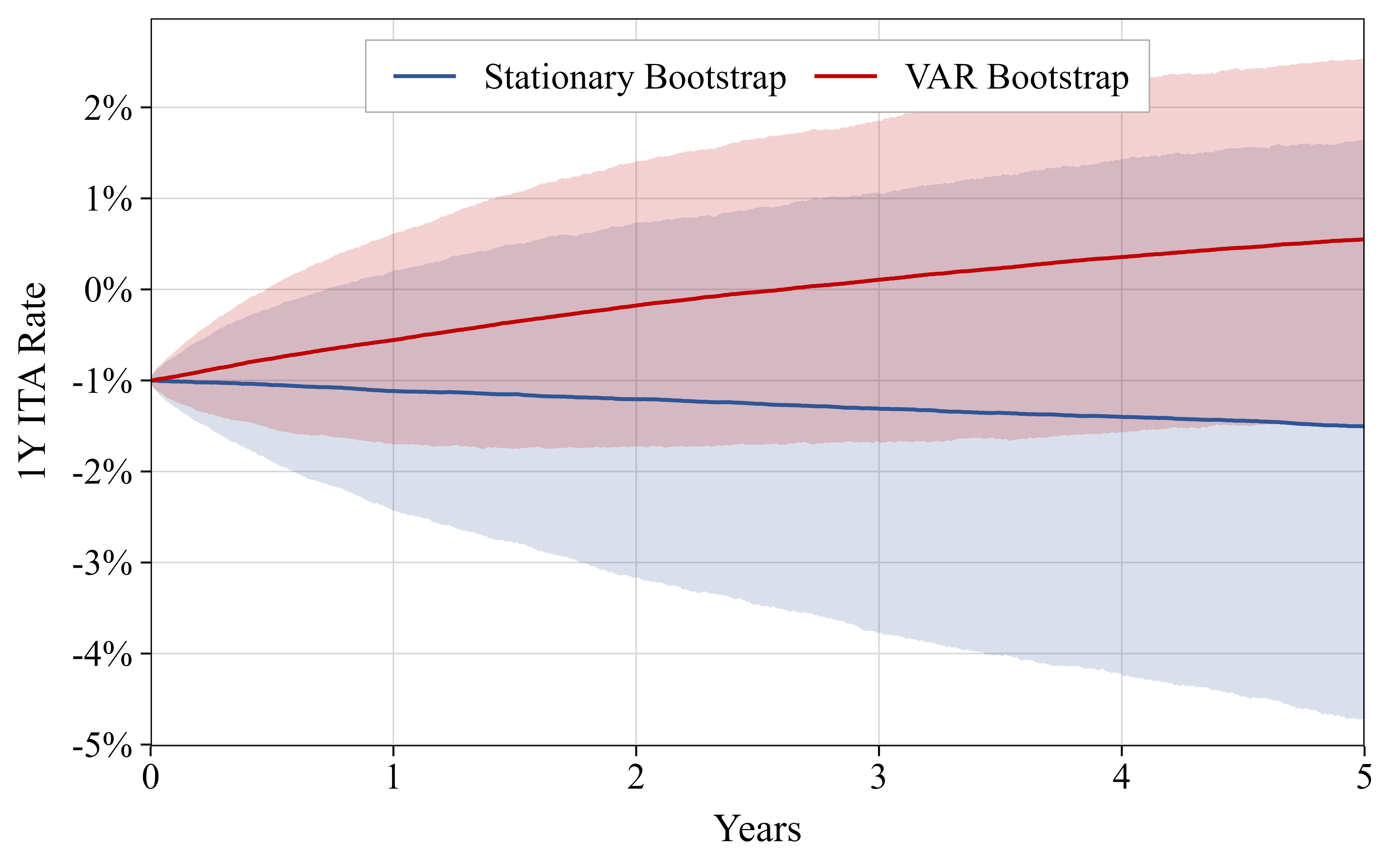}
    \end{minipage}%
    \hfill%
    \begin{minipage}{0.499\textwidth}
        \centering
        \includegraphics[width=\linewidth]{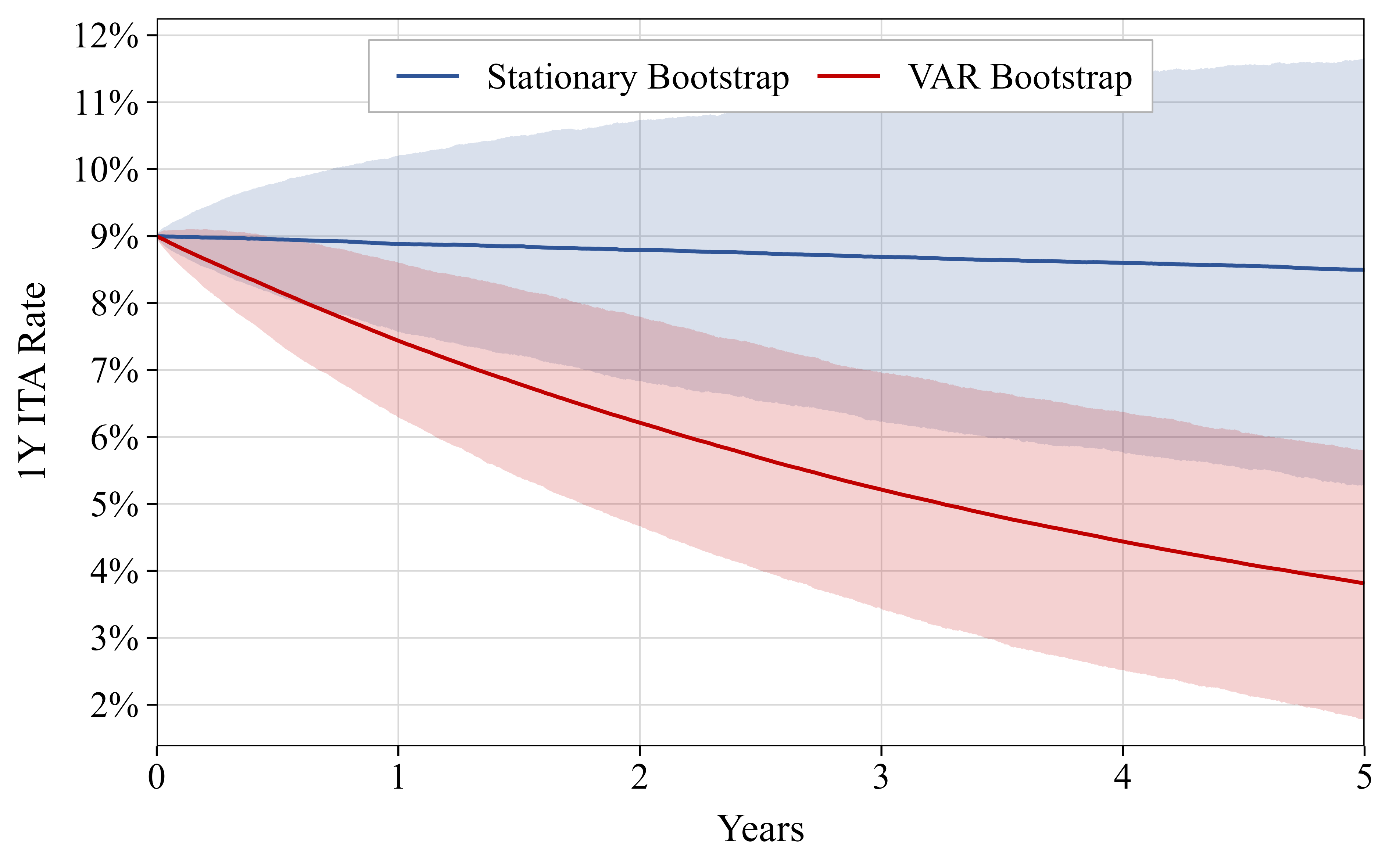}
    \end{minipage}

\caption{Comparison between Stationary Bootstrap and VAR-Bootstrap for the 1-year Italian rate. The trajectories are generated starting from two different initial values, in order to highlight the presence of mean reversion. For each methodology, the mean and the 10\% and 90\% quantiles of the simulated trajectories are reported.}
    \label{fig:var_vs_old_1y}
\end{figure}

\begin{figure}[htbp]
    \centering

    \begin{subfigure}[b]{0.48\textwidth}
        \centering
        \includegraphics[width=\linewidth]{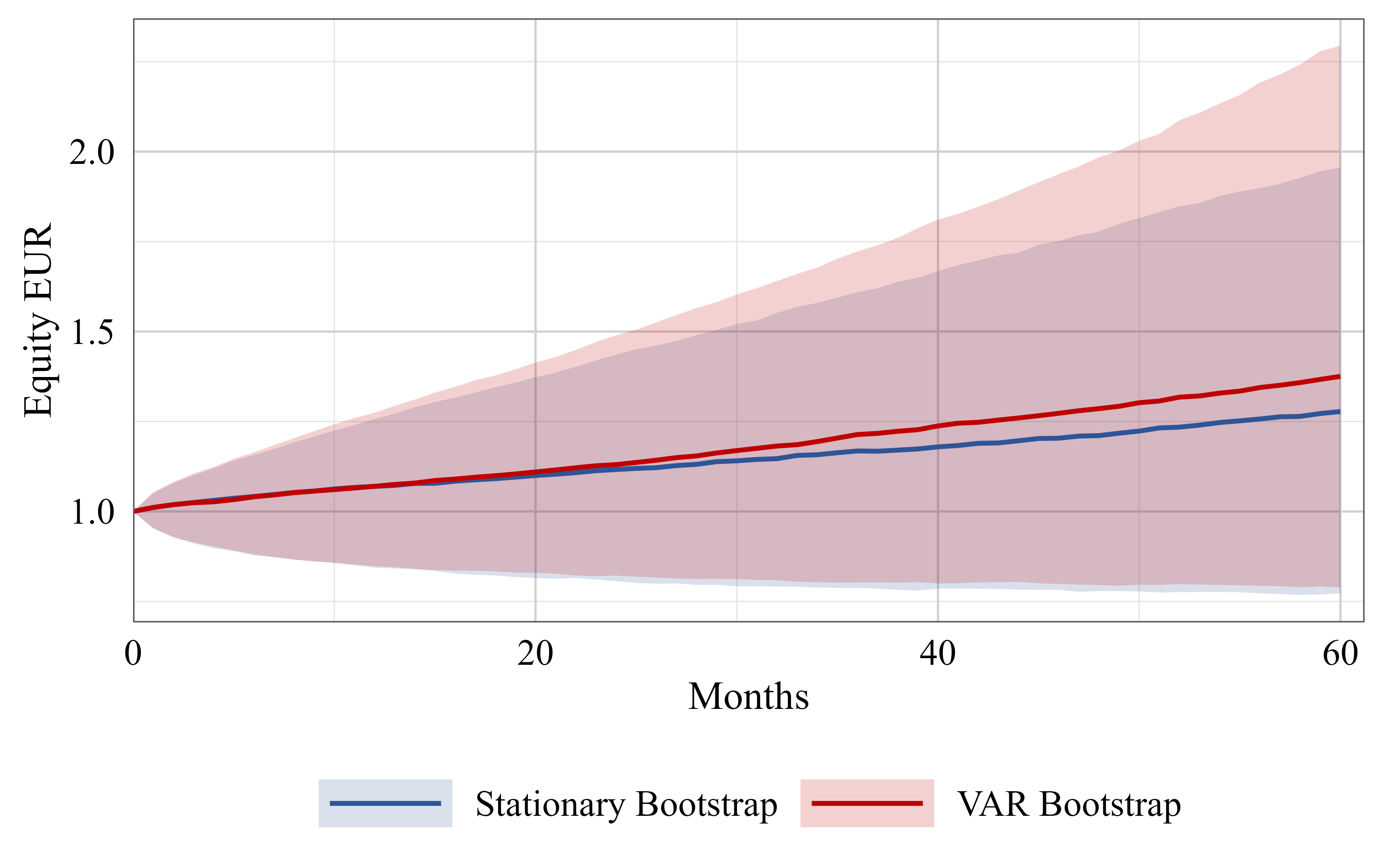}
        \caption{}
        \label{fig:var_vs_old_test2_equity}
    \end{subfigure}

    \vspace{0.08cm}

    \begin{subfigure}[b]{0.48\textwidth}
        \centering
        \includegraphics[width=\linewidth]{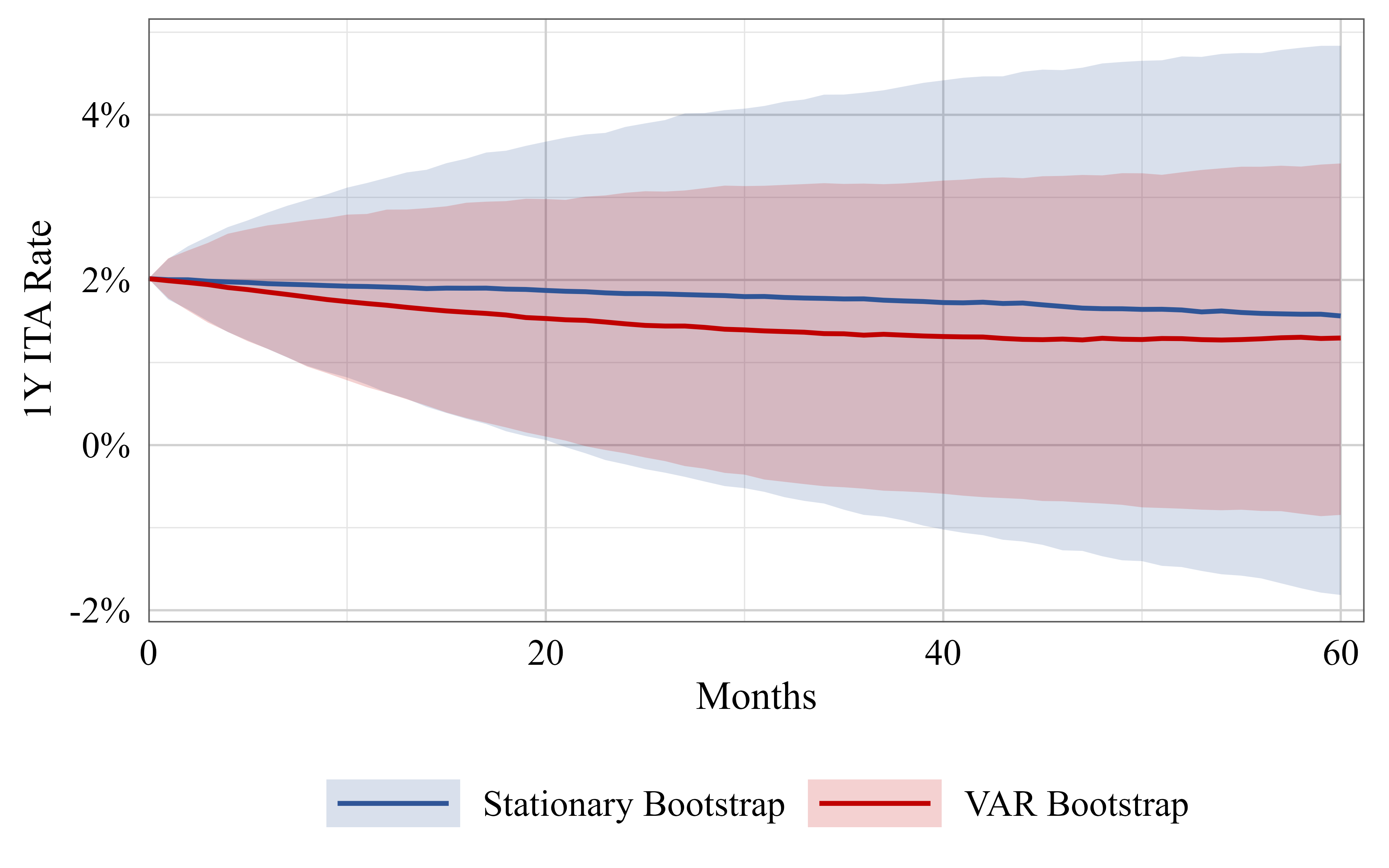}
         \caption{}
        \label{fig:var_vs_old_test2_rate}
    \end{subfigure}

    \vspace{0.08cm}

    \begin{subfigure}[b]{0.48\textwidth}
        \centering
        \includegraphics[width=\linewidth]{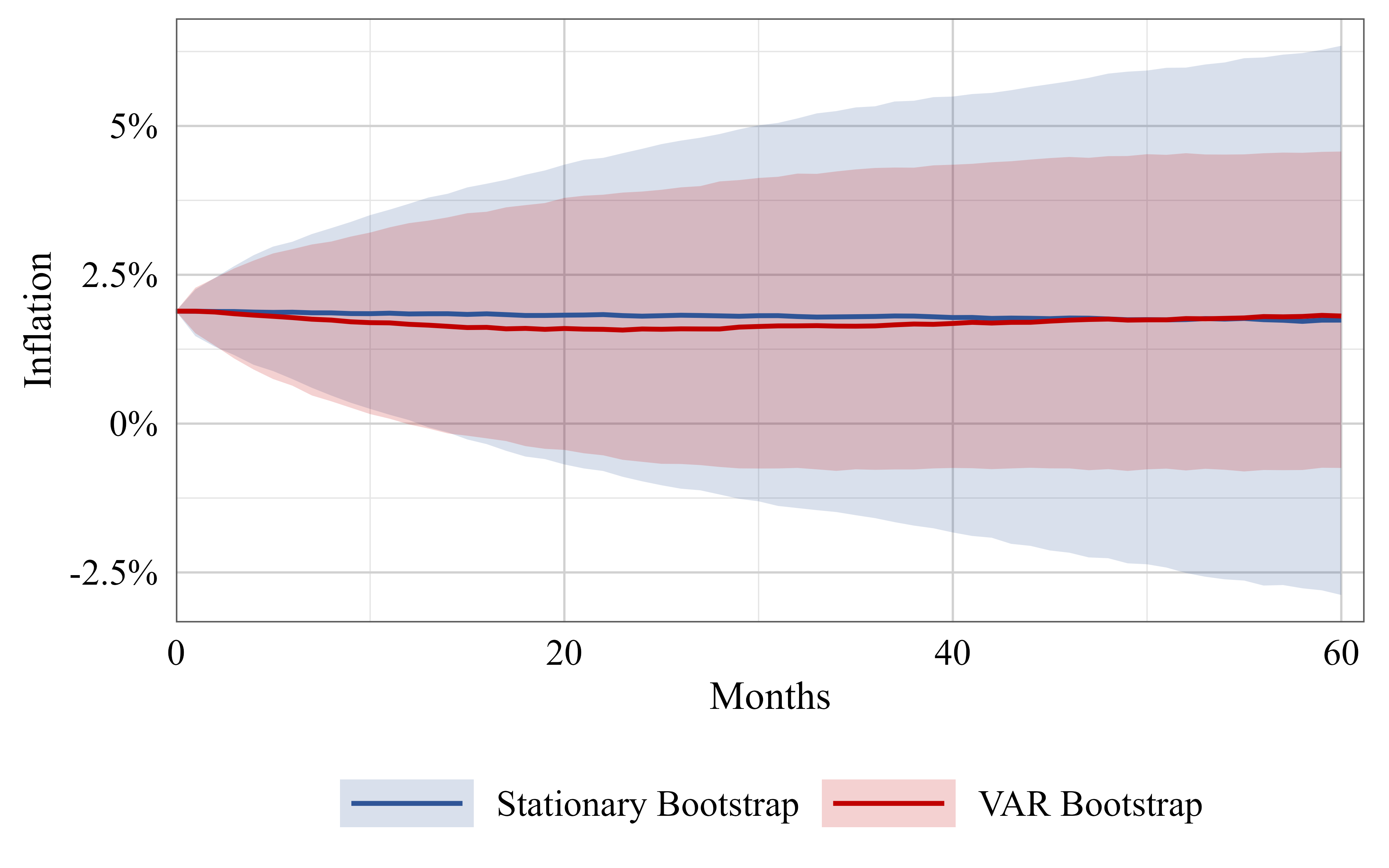}
        \caption{}
        \label{fig:var_vs_old_test2_inflation}
    \end{subfigure}

    \caption{Comparison between Stationary Bootstrap and VAR-Bootstrap for Equity EUR (a), the 1-year Italian rate (b) and inflation (c). Panels (a), (b) and (c) report, respectively, the mean and the 10\% and 90\% quantiles of the simulated trajectories for each methodology.}
    \label{fig:var_vs_old_test2}
\end{figure}

Figure \ref{fig:var_vs_old_1y} compares the simulated trajectories of the 1-year Italian rate obtained by means of the Stationary Bootstrap and the VAR-Bootstrap, considering two different initial values. The figure qualitatively highlights the different behaviour of the two procedures: whereas the Stationary Bootstrap resamples historical variations without introducing an explicit equilibrium level, the VAR-Bootstrap incorporates an autoregressive structure that induces a progressive pull towards the estimated long-run level.

\begin{table}[H]
    \centering
    \small
    \begin{tabular}{@{}lccc@{}}
        \toprule
        Factor & Original series & Stationary Bootstrap & VAR-Bootstrap \\
        \midrule
        Equity EUR & \textbf{45.82\%} & 45.82\% & 46.93\% \\
        1y Rate    & \textbf{49.22\%} & 49.23\% & 53.30\% \\
        \bottomrule
    \end{tabular}
    \normalsize
    \caption{Percentage of negative increments computed on daily data for Equity EUR and for the 1-year Italian rate. The results obtained by means of the Stationary Bootstrap and the VAR-Bootstrap are compared with the corresponding frequencies observed in the historical sample.}
    \label{tab:drawdown_1}
\end{table}

\begin{table}[H]
    \centering
    \small
    \begin{tabular}{@{}lccc@{}}
        \toprule
        Factor & Original series & Stationary Bootstrap & VAR-Bootstrap \\
        \midrule
        Equity EUR & \textbf{41.86\%} & 41.96\% & 40.14\% \\
        1y Rate    & \textbf{54.15\%} & 54.27\% & 55.44\% \\
        Inflation  & \textbf{49.83\%} & 49.84\% & 49.60\% \\
        \bottomrule
    \end{tabular}
    \normalsize
    \caption{Percentage of negative increments computed on monthly data for Equity EUR, the 1-year Italian rate and inflation. The results obtained by means of the Stationary Bootstrap and the VAR-Bootstrap are compared with the corresponding frequencies observed in the historical sample.}
    \label{tab:drawdown_2_3}
\end{table}

\begin{table}[H]
    \centering
    \small
    \begin{tabular}{@{}lccc@{}}
        \toprule
        Factor & Original series & Stationary Bootstrap & VAR-Bootstrap \\
        \midrule
        Equity EUR & \textbf{-0.0083\%} & -0.0083\% & -0.0081\% \\
        1y Rate    & \textbf{-0.0316\%} & -0.0316\% & -0.0297\% \\
        \bottomrule
    \end{tabular}
    \normalsize
    \caption{Mean value computed on the negative increments of Equity EUR and the 1-year Italian rate, using daily data. The results obtained by means of the Stationary Bootstrap and the VAR-Bootstrap are compared with the corresponding values observed in the historical sample.}
    \label{tab:avg_incrementi_negativi_1}
\end{table}

\begin{table}[H]
    \centering
    \small
    \begin{tabular}{@{}lccc@{}}
        \toprule
        Factor & Original series & Stationary Bootstrap & VAR-Bootstrap \\
        \midrule
        Equity EUR & \textbf{-0.0351\%} & -0.0348\% & -0.0344\% \\
        1y Rate    & \textbf{-0.1669\%} & -0.1672\% & -0.1834\% \\
        Inflation  & \textbf{-0.2460\%} & -0.2452\% & -0.2488\% \\
        \bottomrule
    \end{tabular}
    \normalsize
    \caption{Mean value computed on the negative increments of Equity EUR, the 1-year Italian rate and inflation, using monthly data. The results obtained by means of the Stationary Bootstrap and the VAR-Bootstrap are compared with the corresponding values observed in the historical sample.}
    \label{tab:avg_incrementi_negativi_2_3}
\end{table}

\begin{table}[H]
    \centering
    \small
    \begin{tabular}{@{}lccc@{}}
        \toprule
        Pair & Original series & Stationary Bootstrap & VAR-Bootstrap \\
        \midrule
        Equity EUR -- 1y Rate & \textbf{-5.69\%} & -5.26\% & -5.29\% \\
        \bottomrule
    \end{tabular}
    \normalsize
    \caption{Correlation between Equity EUR and the 1-year Italian rate, computed on daily data. The value obtained by means of the Stationary Bootstrap and the VAR-Bootstrap is compared with the corresponding correlation observed in the historical sample.}
    \label{tab:correlation_1}
\end{table}

\begin{table}[H]
    \centering
    \small
    \begin{tabular}{@{}lccc@{}}
        \toprule
        Pair & Original series & Stationary Bootstrap & VAR-Bootstrap \\
        \midrule
        Equity EUR -- 1y Rate   & \textbf{-1.11\%} & -0.52\% & -0.86\% \\
        Equity EUR -- Inflation & \textbf{5.83\%}  & 5.69\%  & 5.52\% \\
        1y Rate -- Inflation    & \textbf{14.90\%} & 15.76\% & 15.83\% \\
        \bottomrule
    \end{tabular}
    \normalsize
    \caption{Correlations between Equity EUR, the 1-year Italian rate and inflation, computed on monthly data. The results obtained by means of the Stationary Bootstrap and the VAR-Bootstrap are compared with the corresponding correlations observed in the historical sample.}
    \label{tab:correlation_2_3}
\end{table}

\begin{table}[H]
    \centering
    \small
    \begin{tabular}{@{}lS[table-format=1.4]S[table-format=1.4]@{}}
        \toprule
        Factor & {Stationary Bootstrap} & {VAR-Bootstrap} \\
        \midrule
        Equity EUR & 0.0050 & 0.0027 \\
        1y Rate    & 0.0082 & 0.0137 \\
        \bottomrule
    \end{tabular}
    \normalsize
    \caption{Kullback--Leibler divergence between the historical empirical distribution and the simulated distributions, computed on daily data for Equity EUR and for the 1-year Italian rate. The table compares the distributional adherence of the simulations generated by means of the Stationary Bootstrap and the VAR-Bootstrap.}
    \label{tab:kl_base_1}
\end{table}

\begin{table}[H]
    \centering
    \small
    \begin{tabular}{@{}lS[table-format=1.4]S[table-format=1.4]S[table-format=1.4]@{}}
        \toprule
        Factor & {Stationary Bootstrap} & {VAR-Bootstrap}\\
        \midrule
        Equity EUR & 0.0005 & 0.0097 \\
        1y Rate    & 0.0345 & 0.2490 \\
        Inflation  & 0.0036 & 0.0118 \\
        \bottomrule
    \end{tabular}
    \normalsize
    \caption{Kullback--Leibler divergence between the historical empirical distribution and the simulated distributions, computed on monthly data for Equity EUR, the 1-year Italian rate and inflation. The table compares the distributional adherence of the simulations generated by means of the Stationary Bootstrap and the VAR-Bootstrap.}
    \label{tab:kl_base_2_3}
\end{table}

Overall, the metrics relating to the negative increments and to the correlations turn out to be generally close to the corresponding historical values for the various methodologies. The Kullback--Leibler divergence favours the Stationary Bootstrap, consistently with its purely empirical nature. The VAR-Bootstrap exhibits a lower adherence to the marginal distributions, especially for rate and inflation, but allows \emph{mean reversion} dynamics to be introduced.

\section{Analysis of the Yield Curve}
\label{sec:results_curve}

The analysis is then extended to the simulation of the entire term structure of Italian interest rates. With respect to the base case, in which the comparison was limited to a single representative rate, in this subsection a set of maturities is considered jointly, so as to assess not only the adherence of the marginal distributions, but also the consistency of the simulated curves along the maturity dimension.

The maturities considered are 3 months, 6 months and the annual maturities from 1 to 10 years. The simulations are carried out on both monthly and daily data, comparing three methodologies: the Stationary Bootstrap, the VAR-Bootstrap applied directly to the levels of the yield curve, and the Nelson-Siegel VAR-Bootstrap, in which the autoregressive dynamics is estimated on the Nelson-Siegel latent factors and the curve is subsequently reconstructed from such factors.

\begin{table}[H]
    \centering
    \small
    \resizebox{\textwidth}{!}{%
    \begin{tabular}{@{}l*{6}{S[table-format=1.4]}@{}}
        \toprule
        & \multicolumn{3}{c}{Monthly data} & \multicolumn{3}{c}{Daily data} \\
        \cmidrule(lr){2-4}\cmidrule(l){5-7}
        Maturity
        & {Stationary Bootstrap}
        & {VAR-Bootstrap}
        & {Nelson-Siegel VAR-Bootstrap}
        & {Stationary Bootstrap}
        & {VAR-Bootstrap}
        & {Nelson-Siegel VAR-Bootstrap} \\
        \midrule
        3m  & 0.0006 & 0.3275 & 0.1962 & 0.0032 & 0.0149 & 0.0723 \\
        6m  & 0.0006 & 0.2111 & 0.1780 & 0.0003 & 0.0039 & 0.0233 \\
        1y  & 0.0004 & 0.0938 & 0.1734 & 0.0003 & 0.0006 & 0.1968 \\
        2y  & 0.0005 & 0.1120 & 0.0461 & 0.0006 & 0.0011 & 0.0133 \\
        3y  & 0.0004 & 0.1266 & 0.0350 & 0.0005 & 0.0007 & 0.0014 \\
        4y  & 0.0005 & 0.0590 & 0.0945 & 0.0007 & 0.0009 & 0.0310 \\
        5y  & 0.0003 & 0.0819 & 0.0396 & 0.0004 & 0.0014 & 0.1182 \\
        6y  & 0.0004 & 0.1009 & 0.0517 & 0.0004 & 0.0006 & 0.0195 \\
        7y  & 0.0003 & 0.1031 & 0.0212 & 0.0003 & 0.0007 & 0.0014 \\
        8y  & 0.0003 & 0.0546 & 0.0160 & 0.0002 & 0.0003 & 0.0814 \\
        9y  & 0.0004 & 0.0238 & 0.0094 & 0.0001 & 0.0004 & 0.0016 \\
        10y & 0.0004 & 0.1159 & 0.0873 & 0.0001 & 0.0002 & 0.0137 \\
        \bottomrule
    \end{tabular}%
    }
    \normalsize
    \caption{Kullback--Leibler divergence between the historical empirical distribution and the simulated distributions of Italian rates, computed separately for each maturity. The table compares the distributional adherence of the simulations generated by means of the Stationary Bootstrap, the VAR-Bootstrap and the Nelson-Siegel VAR-Bootstrap, on both monthly and daily data.}
    \label{tab:kl_rates_results}
\end{table}

\begin{figure}[]
    \centering

    \includegraphics[width=0.72\textwidth]{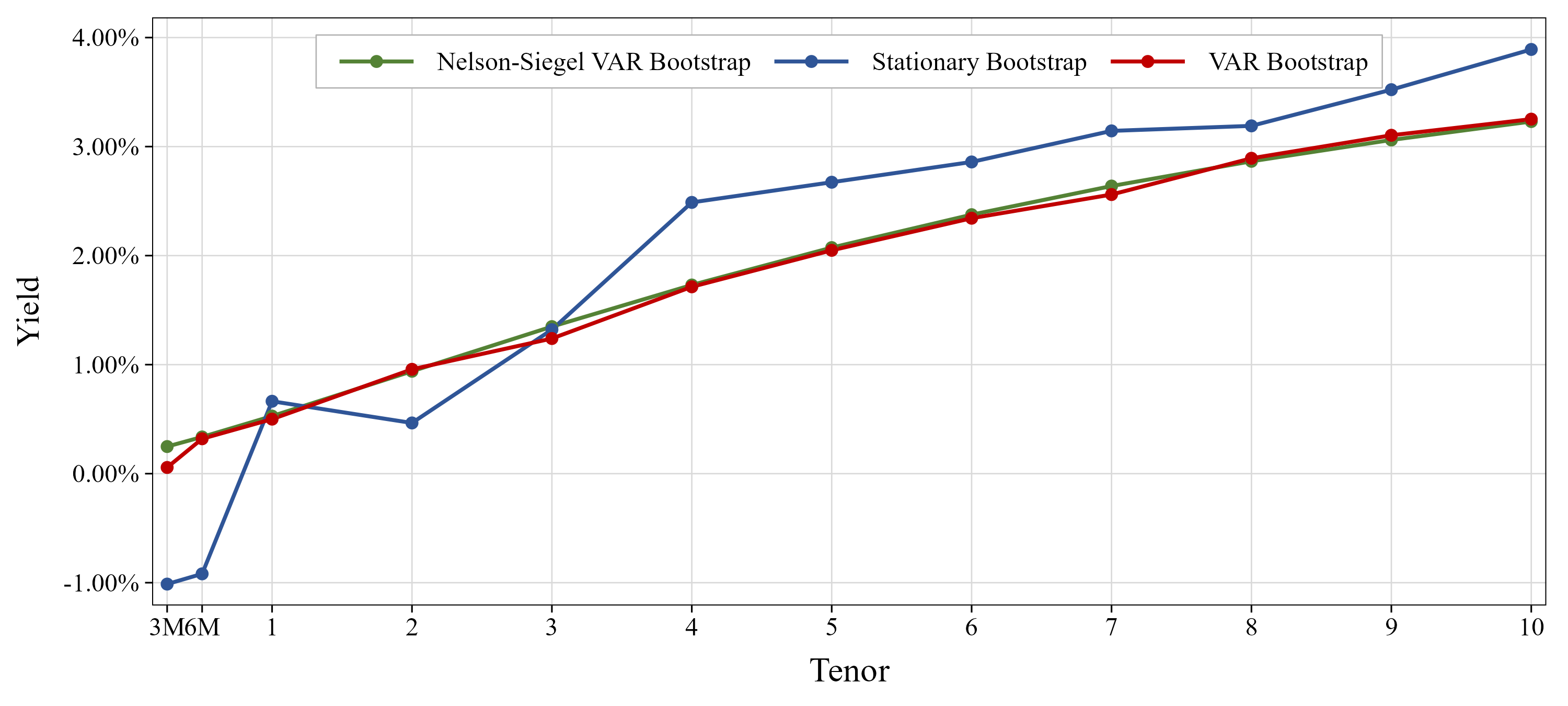}

    \vspace{0.15cm}

    \includegraphics[width=0.72\textwidth]{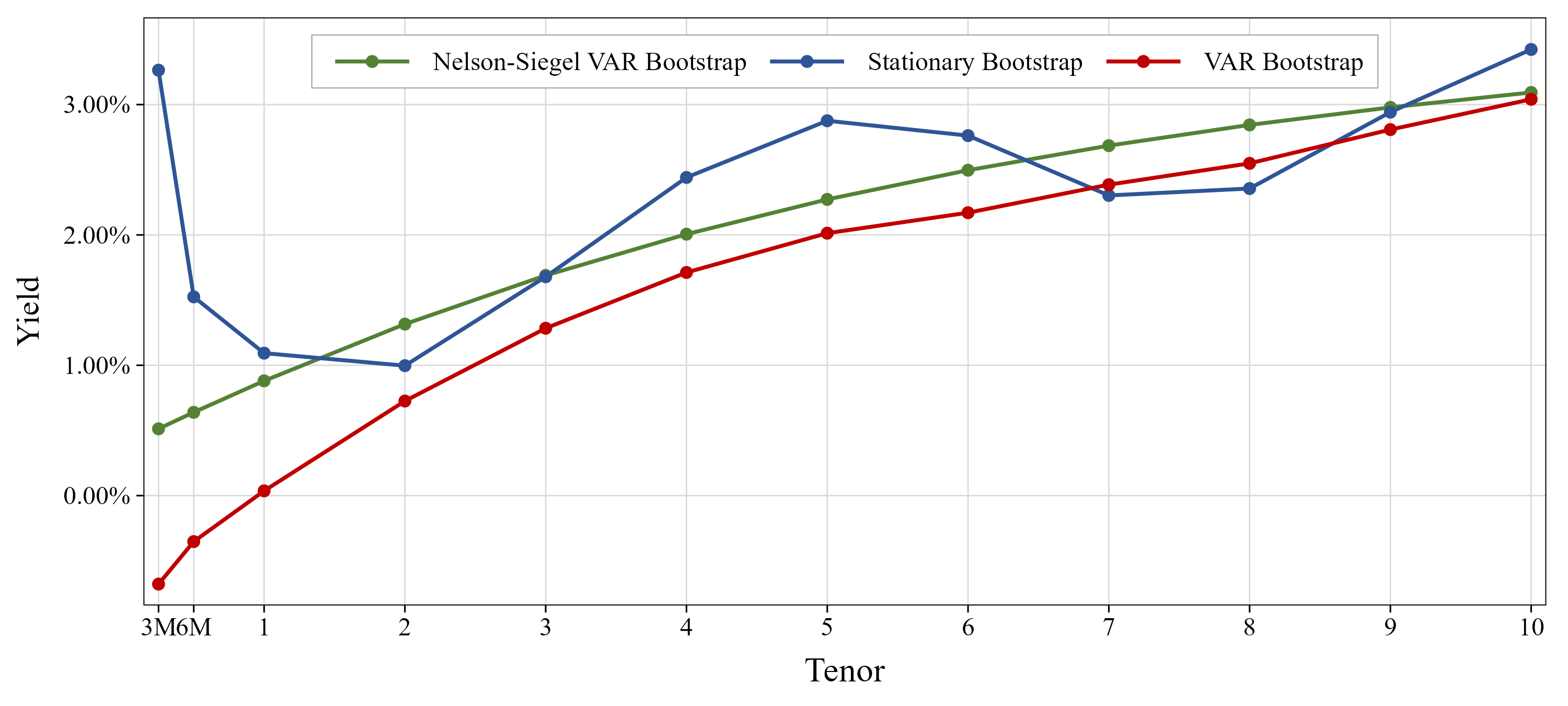}

    \caption{Examples of simulated Italian yield curves at the five-year projection date. The two panels compare, for two simulated scenarios, the profiles generated by means of the Stationary Bootstrap, the VAR-Bootstrap and the Nelson-Siegel VAR-Bootstrap, highlighting the different regularity of the curves along the maturities.}
    \label{fig:curve_5y_scenari}
\end{figure}

Table \ref{tab:kl_rates_results} reports, for each maturity, the Kullback--Leibler divergence between the historical empirical distribution and the simulated distributions. Consistently with the empirical nature of the procedure, the Stationary Bootstrap generally exhibits lower values of the metric, since the direct resampling from historical observations tends to preserve the marginal distributions with greater accuracy. The models based on autoregressive specifications, by contrast, show higher KL values, especially at certain maturities, reflecting the dynamic constraint imposed by the parametric structure.

This result must, however, be interpreted together with the regularity of the simulated curves. Figure \ref{fig:curve_5y_scenari} shows examples of curves generated over a 5-year horizon and highlights how the VAR-Bootstrap and Nelson-Siegel VAR-Bootstrap models, although less adherent to the marginal distributions, tend to produce more regular profiles along the maturities. In particular, the representation by means of Nelson-Siegel factors makes it possible to impose a more parsimonious structure on the curve, reducing the possibility of irregular movements between adjacent maturities.

A further aspect to be considered in the evaluation of the Nelson-Siegel VAR-Bootstrap concerns the fitting error of the factor representation of the curve. This error is described by the residuals \(\varepsilon_t(\tau)\), defined in equation \eqref{eq:tassi_NS}, and measures the portion of the observed rates that is not explained by the Nelson-Siegel parameterisation.

Considered jointly along the temporal dimension and along the various maturities, the residuals turn out to be overall centred around zero and present an approximately symmetric, though non-Gaussian, distribution. The high kurtosis signals a greater concentration around the mean and heavier tails with respect to the normal distribution.

The analysis by individual maturity nonetheless reveals greater heterogeneity along the curve. In particular, the distributions of the residuals associated with the various maturities do not turn out to be perfectly superimposable: for some maturities the mean deviates from zero and limited elements of asymmetry are observed. This evidence suggests that the fitting error is not uniform across the entire term structure and that the model may exhibit, in specific areas of the curve, slight systematic tendencies towards underestimation or overestimation of the observed rates. The overall magnitude of the deviations nevertheless remains limited.

\begin{table}[H]
    \centering
    \small
    \begin{tabular}{@{}lS[table-format=-1.4]S[table-format=-1.4]@{}}
        \toprule
        Maturity & {Mean} & {Skewness} \\
        \midrule
        \textbf{Curve} & \bfseries 0.0000 & \bfseries -0.2069 \\
        \midrule
        3m  & -0.0511 & -1.2763 \\
        1y  & 0.0138  & 0.3917 \\
        5y  & -0.0093 & 0.2742 \\
        10y & 0.0497  & 1.2722 \\
        \bottomrule
    \end{tabular}
    \normalsize
    \caption{Descriptive statistics of the residuals \(\varepsilon_t(\tau)\) of the Nelson-Siegel parameterisation. The table reports the mean and skewness computed both on the overall set of residuals, across maturities and over the time dimension, and on a number of representative maturities.}
    \label{tab:NS_res}
\end{table}

%% file: Chapters/9.Conclusions.tex
\section{Conclusions}
\label{sec:conclusions}

The objective of this work was to compare different simulation methodologies for financial and macroeconomic risk factors, with particular attention to the trade-off between statistical adherence to the historical data and economic-financial plausibility of the simulated scenarios.

The empirical results show that the Stationary Bootstrap provides the highest degree of adherence to the historical marginal distributions. This outcome is consistent with the construction of the method, which is based on the direct resampling of historical observations or historical increments. The values of the Kullback--Leibler divergence are generally lower for the Stationary Bootstrap, reflecting its ability to preserve the empirical distribution of the data.

However, this greater distributional adherence comes at the cost of a limited economic structure. Since the Stationary Bootstrap relies on the recombination of historical observations rather than on an explicit structural mechanism, the simulated trajectories may reproduce historical features at the aggregate level without necessarily being individually plausible from a dynamic point of view. The VAR-Bootstrap addresses this limitation by introducing an autoregressive structure in the simulated dynamics. The results show that this approach is able to generate trajectories characterised by a progressive pull towards the estimated long-run levels, especially for the rate component. This property improves the economic plausibility of the scenarios, even though it generally reduces the adherence of the simulated marginal distributions to the historical empirical ones. 

The extension of the analysis to the entire Italian yield curve further highlights the different nature of the methodologies considered. The Stationary Bootstrap remains the best-performing approach in terms of marginal distributional adherence across maturities, as measured by the Kullback--Leibler divergence. Nevertheless, the simulated curves generated through purely empirical resampling may display irregular movements along the maturity dimension. By contrast, the VAR-Bootstrap and the Nelson-Siegel VAR-Bootstrap introduce a more structured representation of the term structure, leading to smoother and more coherent yield curves.

Overall, the results indicate that no methodology dominates the others in all dimensions. The Stationary Bootstrap is preferable when the primary objective is the preservation of the historical empirical distribution. The VAR-Bootstrap is more appropriate when the simulation must incorporate dynamic properties such as mean reversion and dependence across factors. The Nelson-Siegel VAR-Bootstrap becomes particularly relevant when the objective is the joint simulation of the yield curve, since it combines autoregressive dynamics with a parsimonious and regular representation of the term structure.

%% file: Chapters/A.Appendix.tex
\section{Appendix: Consistency in Distribution}
\label{sec:consistency_distribution}

This section clarifies an important difference between purely nonparametric resampling methods and semiparametric bootstrap procedures based on autoregressive dynamics. The key point is that, while a nonparametric bootstrap resamples directly from the empirical distribution of the observed data, an AR-Bootstrap resamples innovations and then propagates them through a dynamic model. As a consequence, the simulated observations are not, in general, distributed as the original observations.

Consider a time series of observed values \(\{r_1,\ldots,r_T\}\). A traditional nonparametric approach, such as the stationary bootstrap, generates simulated observations by resampling from the empirical distribution of the data. In the simple case of independent observations, this implies that, asymptotically,
\[
    r_t \overset{\mathrm{iid}}{\sim} d
    \quad \Longrightarrow \quad
    r_t^* \overset{\mathrm{iid}}{\sim} d,
\]
where \(d\) denotes the empirical distribution of the observed sample under a suitable probability measure. In this sense, the bootstrap observations preserve the marginal distribution of the original data.

Consider now, for simplicity, an AR-Bootstrap engine of the form
\[
    r_t = a_0 + a_1 r_{t-1} + \eta_t,
\]
which generates simulated time series according to
\[
    r_t^* = a_0 + a_1 r_{t-1}^* + \eta_t^*.
\]
In this case, the bootstrap is applied to the residuals \(\eta_t\), rather than directly to the observations \(r_t\). Therefore, even if the residuals are resampled from their empirical distribution, the simulated values \(r_t^*\) are obtained after applying the autoregressive recursion. In general,
\[
    \boxed{r_t^* \not\sim r_t.}
\]
The same mechanism arises in VAR-Bootstrap models, and even more strongly so: in that case, each simulated component depends not only on its own lagged values, but also on the lagged values of the other variables in the system. Hence, the dynamic propagation of resampled shocks further modifies the marginal distributions of the simulated series.

To illustrate the point, consider the simple observed path
\[
    (r_0,r_1,r_2) = (4,2.2,1).
\]
Here \(r_0=4\) is assumed to be known and fixed. It represents the starting point of the time series and is used only as an anchoring initial condition for the simulation. It is not part of the resampling step. The bootstrap sampling is instead performed only on the observed values \(r_1\) and \(r_2\).

Under a purely nonparametric bootstrap, at each simulated date one draws directly from the empirical distribution of \(\{r_1,r_2\}\). Hence, at \(t=1\),
\[
    r_1^* =
    \begin{cases}
        2.2, & \text{with probability } 1/2,\\
        1,   & \text{with probability } 1/2.
    \end{cases}
\]
At the following step, the same empirical distribution is used again, with replacement. Therefore,
\[
    r_2^* =
    \begin{cases}
        2.2, & \text{with probability } 1/2,\\
        1,   & \text{with probability } 1/2.
    \end{cases}
\]
Thus, for each simulated date, the marginal distribution of \(r_t^*\) coincides with the empirical distribution of the sampled observations. In this example, $\mathbb{P}(r_t^*=2.2)
    =
    \mathbb{P}(r_t^*=1)
    =
    \frac{1}{2}.$

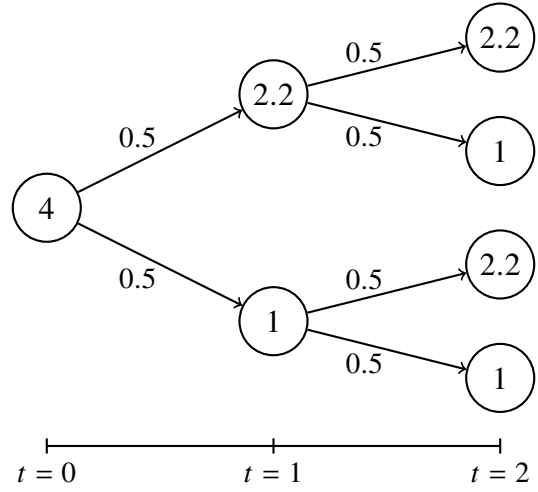
\begin{figure}[H]
    \centering
    \begin{tikzpicture}[
        node distance=2.8cm,
        state/.style={circle, draw=arcaBlue, thick, minimum size=0.9cm, inner sep=1pt},
        arrow/.style={->, thick, arcaBlue},
        prob/.style={font=\small, fill=white, inner sep=1pt},
        time/.style={font=\small\bfseries, text=arcaBlue}
    ]
        \draw[arcaBlack, thick] (0,-3.15) -- (6,-3.15);
        \foreach \xcoord/\label in {0/{$t=0$},3/{$t=1$},6/{$t=2$}} {
            \draw[arcaBlack, thick] (\xcoord,-3.25) -- (\xcoord,-3.05);
            \node[time] at (\xcoord,-3.5) {\label};
        }

        \node[state] (x) at (0,0) {$4$};
        \node[state] (y1) at (3,1.5) {$2.2$};
        \node[state] (y2) at (3,-1.5) {$1$};
        \node[state] (z1) at (6,2.25) {$2.2$};
        \node[state] (z2) at (6,0.75) {$1$};
        \node[state] (z3) at (6,-0.75) {$2.2$};
        \node[state] (z4) at (6,-2.25) {$1$};

        \draw[arrow] (x) -- node[prob, above left] {$0.5$} (y1);
        \draw[arrow] (x) -- node[prob, below left] {$0.5$} (y2);
        \draw[arrow] (y1) -- node[prob, above left] {$0.5$} (z1);
        \draw[arrow] (y1) -- node[prob, below left] {$0.5$} (z2);
        \draw[arrow] (y2) -- node[prob, above left] {$0.5$} (z3);
        \draw[arrow] (y2) -- node[prob, below left] {$0.5$} (z4);
    \end{tikzpicture}
    \caption{Nonparametric Bootstrap Illustration.}
    \label{fig:branching_scheme}
\end{figure} Let us now consider an AR-Bootstrap. The first step is to estimate the underlying autoregressive model. Suppose, for example, that the fitted parameters are
\[
    a_0=0,
    \qquad
    a_1=0.5.
\]
The corresponding residuals are
\[
    \eta_1 = 2.2 - 0.5 \times 4 = 0.2,
    \qquad
    \eta_2 = 1 - 0.5 \times 2.2 = -0.1.
\]
The simulation starts from the fixed initial value \(r_0^*=r_0=4\), and the residuals are resampled from \(\{0.2,-0.1\}\). Hence,
\[
    r_1^* = 0.5 \times 4 + \eta_1^*
          = 2 + \eta_1^*,
\]
so \(r_1^*\) can be either \(2.2\) or \(1.9\), each with probability \(1/2\). Already at this stage, the simulated marginal distribution differs from the empirical distribution of the observations \(\{2.2,1\}\). The difference becomes more evident at the following step, since
\[
    r_2^* = 0.5 r_1^* + \eta_2^*.
\]
The value of \(r_2^*\) therefore depends on the previously simulated value \(r_1^*\), not only on the newly resampled residual. As a result, the possible values are $\{1.3,1.0,1.15,0.85\},$ rather than the original empirical values \(\{2.2,1\}\). This illustrates how the autoregressive recursion transforms the empirical distribution of the resampled residuals into a different distribution for the simulated observations.

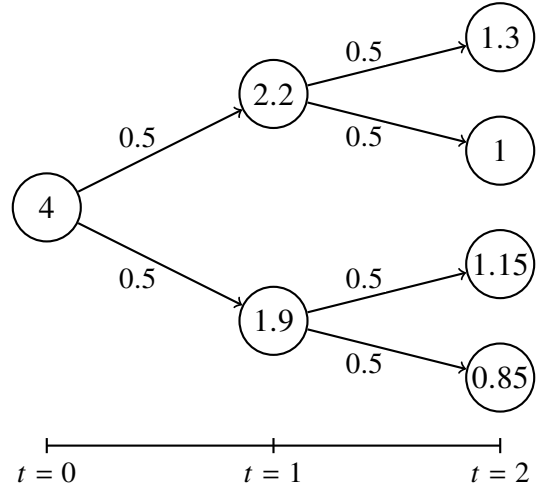
\begin{figure}[H]
    \centering
    \begin{tikzpicture}[
        node distance=2.8cm,
        state/.style={circle, draw=arcaBlue, thick, minimum size=0.9cm, inner sep=1pt},
        arrow/.style={->, thick, arcaBlue},
        prob/.style={font=\small, fill=white, inner sep=1pt},
        time/.style={font=\small\bfseries, text=arcaBlue}
    ]

    \draw[arcaBlack, thick] (0,-3.15) -- (6,-3.15);
        \foreach \xcoord/\label in {0/{$t=0$},3/{$t=1$},6/{$t=2$}} {
            \draw[arcaBlack, thick] (\xcoord,-3.25) -- (\xcoord,-3.05);
            \node[time] at (\xcoord,-3.5) {\label};
        }

        \node[state] (x) at (0,0) {$4$};
        \node[state] (y1) at (3,1.5) {$2.2$};
        \node[state] (y2) at (3,-1.5) {$1.9$};
        \node[state] (z1) at (6,2.25) {$1.3$};
        \node[state] (z2) at (6,0.75) {$1$};
        \node[state] (z3) at (6,-0.75) {$1.15$};
        \node[state] (z4) at (6,-2.25) {$0.85$};

        \draw[arrow] (x) -- node[prob, above left] {$0.5$} (y1);
        \draw[arrow] (x) -- node[prob, below left] {$0.5$} (y2);
        \draw[arrow] (y1) -- node[prob, above left] {$0.5$} (z1);
        \draw[arrow] (y1) -- node[prob, below left] {$0.5$} (z2);
        \draw[arrow] (y2) -- node[prob, above left] {$0.5$} (z3);
        \draw[arrow] (y2) -- node[prob, below left] {$0.5$} (z4);
    \end{tikzpicture}
    \caption{Semiparametric Bootstrap Illustration.}
    \label{fig:semiparametric_branching_scheme}
\end{figure} The implication is explicit: the AR-Bootstrap does not preserve the empirical distribution of the observed values. In the non-parametric case, at each simulated date the support of the distribution is exactly the empirical support \(\{2.2,1\}\), with probability \(1/2\) assigned to each value. In the AR-Bootstrap case, instead, the support is transformed by the autoregressive recursion. At \(t=1\), the simulated support is already \(\{2.2,1.9\}\), and at \(t=2\) it becomes $\{1.3,1.0,1.15,0.85\}.$

This example shows that, in an AR-Bootstrap, the resampled residuals are filtered through the autoregressive structure. The simulated observations are therefore generated by combining sampled shocks with previously simulated values. This recursive dependence changes the marginal distribution of the simulated series. The same argument applies to VAR-Bootstrap models, where the effect is even more pronounced because each variable is dynamically linked to the past values of all variables included in the system.
\newpage
\section{Appendix: Absence of Arbitrage}
\label{sec:aoa}

The simulation methodologies considered do not guarantee, by construction, that the generated yield curves are compatible with no-arbitrage conditions. For this reason, the financial consistency of the simulated curves is verified \emph{ex post} through a criterion based on the implied discount factors. The analysis is limited to scenarios in which the entire yield curve turns out to be non-negative, so that the monotonicity of the discount factors can be interpreted directly as a condition of financial consistency.

Let \(B(0,T)\) denote the price at time \(0\) of a zero-coupon bond paying one monetary unit at maturity \(T\). Assuming continuous compounding, the discount factor relative to maturity \(\tau\) is given by
\begin{equation}
    B(0,\tau)
    =
    e^{-\tau y(\tau)}.
\end{equation}

In the absence of arbitrage, and under the assumption of non-negative rates, the discount factors must be non-increasing with respect to maturity. In other terms, a certain payment equal to \(1\) available at a more distant date cannot have today a price higher than that of the same payment available at a closer date. Therefore, for two maturities \(T_1<T_2\), the following must hold:
\begin{equation}
    1
    =
    B(0,0)
    \geq
    B(0,T_1)
    \geq
    B(0,T_2)
    \geq
    0.
    \label{eq:aoa_curva}
\end{equation}

The reason is as follows. If one were to observe \(B(0,T_1)<B(0,T_2)\), the security with the shorter maturity would be less expensive than the one with the longer maturity, even though both pay the same unit amount. It would therefore be possible to buy the zero-coupon bond with maturity \(T_1\) and to short-sell the one with maturity \(T_2\), obtaining at the initial time a positive net inflow equal to \(B(0,T_2)-B(0,T_1)\). At maturity \(T_1\), the security purchased pays back \(1\). Since only scenarios with non-negative rates are considered, this amount can be held or reinvested up to \(T_2\) without losing value, and is therefore sufficient to cover the payment due on the short position. The strategy thus generates a positive initial inflow without requiring any net future outlay, constituting an arbitrage opportunity.

Operationally, the analysis is carried out in two stages. First, the percentage of simulated curves that exhibit non-negative rates at all maturities considered is computed. Subsequently, within this subset, the share of curves that violate the monotonicity condition of the discount factors is determined. The results are reported in Table \ref{tab:aoa_curva}.

\begin{table}[H]
    \centering
    \small
    \begin{tabular}{@{}lcc@{}}
        \toprule
        Method & \makecell{Absence of\\negative rates} & \makecell{Presence of\\arbitrage} \\
        \midrule
        Stationary Bootstrap        & 77.44\% & 21.66\% \\
        VAR-Bootstrap               & 87.84\% & 0.56\% \\
        Nelson-Siegel VAR-Bootstrap & 88.51\% & 0.24\% \\
        \bottomrule
    \end{tabular}
    \normalsize
    \caption{Share of simulated curves with non-negative rates at all maturities and percentage of violations of the monotonicity of the discount factors within this subset.}
    \label{tab:aoa_curva}
\end{table}

The results highlight that the Stationary Bootstrap generates a significantly higher share of curves with violations of the no-arbitrage condition. This is consistent with the nature of the method, which resamples historical observations directly and preserves the marginal distributions, but does not impose constraints on the overall shape of the curve. The VAR-Bootstrap and Nelson-Siegel VAR-Bootstrap models, by contrast, significantly reduce such violations, since they introduce a common structure in the dynamics of the maturities. In particular, the Nelson-Siegel factor representation favours the generation of more regular profiles along the curve, improving its financial consistency.